\documentclass[%
reprint,
superscriptaddress,
 amsmath,amssymb,
 aps,
 prx,
floatfix,
]{revtex4-2}

\usepackage[english]{babel}

\usepackage[hidelinks]{hyperref}
\usepackage{graphicx}
\usepackage{dcolumn}
\usepackage{bm}

\usepackage{comment}
\usepackage{physics}
\usepackage{cleveref}
\usepackage{xfrac}
\usepackage{xcolor}
\usepackage{booktabs}
\usepackage{tabularx}
\usepackage{threeparttable}

\usepackage{float}

\begin{document}

\preprint{APS/123-QED}

\title{An optical-fibre-integrated buffer for packet-switched quantum networks}


\author{Daniel~Spegel-Lexne}
\email{daniel.spegel-lexne@liu.se}
\affiliation{Institutionen för Systemteknik, Linköpings Universitet, 581 83, Linköping, Sweden}

\author{Joakim~Argillander}
\affiliation{Institutionen för Systemteknik, Linköpings Universitet, 581 83, Linköping, Sweden}

\author{Martin~Clason}
\affiliation{Institutionen för Systemteknik, Linköpings Universitet, 581 83, Linköping, Sweden}

\author{Åsa~Claesson}
\affiliation{Fibrelab, RISE-Research Institutes of Sweden, Fibrevägen 2-6, 82450, Hudiksvall, Sweden}

\author{Kenny~Hey~Tow}
\affiliation{Fibre Optics, RISE-Research Institutes of Sweden, Electrum 236, 16440, Kista, Sweden}

\author{Gustavo~Lima}
\affiliation{Departamento de F\'{\i}sica, Universidad de Concepci\'on,
             160-C Concepci\'on, Chile}
\affiliation{Millennium Institute for Research in Optics,
             Universidad de Concepci\'on, 160-C Concepci\'on, Chile}

\author{João~M.~B.~Pereira}
\affiliation{Fibre Optics, RISE-Research Institutes of Sweden, Electrum 236, 16440, Kista, Sweden}

\author{Guilherme~B.~Xavier}
\email{guilherme.b.xavier@liu.se}
\affiliation{Institutionen för Systemteknik, Linköpings Universitet, 581 83, Linköping, Sweden}

\date{\today}

\begin{abstract}

Packet-switched quantum networks require buffers that can delay qubit payloads while routing information is read out in real time. Previous approaches have not provided this functionality in a fully fibre-integrated architecture compatible with telecom infrastructure. Here we demonstrate an optical-fibre-integrated buffer, based on a recirculating loop and a fibre storage line, in which the storage time of a polarisation-encoded qubit payload is determined by readout of an attached packet header. The key component behind this achievement is an ultra-low-loss poled fibre phase modulator, which provides fast, polarisation-insensitive switching directly in fibre and allows header and payload to be processed within the buffer. We demonstrate storage and retrieval of polarisation-encoded qubit payloads for storage times up to 47 $\mu$s, with an average quantum bit error rate of 1.8\% together with stable operation over several hours. These results establish a practical fibre-based architecture for packet-level quantum network buffering that can easily integrate into the current telecommunication infrastructure opening up new paths for deployment of the quantum internet.

\end{abstract}

\maketitle

\section*{Introduction}

Realizing large-scale quantum networks, often framed as a “quantum internet”, would enable functionalities that are fundamentally out of reach for classical communication networks, including device-independent security and distributed quantum computing \cite{kimble2008quantum,wehner2018quantum}. Essentially, in all architectures, photons in the telecom band are the natural carriers of quantum information across metropolitan and intercity links owing to the low-loss properties of commercial optical fibres in the 1550 nm spectral region \cite{gisin2007quantum, pirandola2020advances}. Quantum key distribution (QKD) is a well-established quantum communication application, having had many demonstrations ranging from foundational experiments to modern, ultra-long-distance and side-channel attack secure versions \cite{bennett1984bb84,ekert1991bell,lo2012mdi,TF2018}, as well as field demonstrations now spanning terrestrial fibre networks and free-space and satellite links \cite{liao2017satelliteqkd,yin2017satelliteentanglement, zhou2024combfield, pittaluga2025deployed}. As these systems evolve from point-to-point demonstrations to multi-user networks, they increasingly require network-layer primitives, such as packet buffering, without degrading the fragile quantum states \cite{wehner2018quantum,pompili2022networkstack}.

Recently, significant effort has been dedicated to developing a framework for packet-switched quantum networks as a natural extension from classical packet-switched communication extensively used over the Internet \cite{DiAdamo2022, Mandil2023, Yoo2024}.  Within a network node, quantum memories \cite{lvovsky2009memory,heshami2016memories,bussieres2013applications} are needed to act as buffers allowing time for header processing belonging to each quantum packet. Typical quantum memory implementations, although nowadays offering excellent performance in terms of storage lifetime or efficiency, have the general disadvantage of being relatively complex and not directly spectrally compatible with optical-fibre telecom infrastructure, as well as requiring cryogenic cooling and bulk optical elements making them challenging to integrate into real-world network nodes \cite{Ortu2022, Duranti2024}. 

Another approach to try to bridge the gap towards more practical quantum memories is based on the use of an optical loop (working as the quantum memory), combined with an optical switch for storage/retrieval \cite{Pittman2002}. This approach has been highly successful with many relevant applications being demonstrated \cite{Makino2016, Kaneda2017, Kaneda2019, Meyer-Scott2022, Hou2023, Weinbrenner2024, Hanamura2025, Pang2023, Bonsma-Fisher2023, Bonsma-Fisher:24}. All these previous approaches relied on bulk optical elements, making it difficult to integrate with fibre-optical components, thus greatly limiting their use for quantum communication and networking applications. Recently, efforts on fibre-based recirculating memories for quantum applications have included a fibre loop with an electro-optical switch with storage periods of several $\mu s$ \cite{FookLee2024}. Other very recent approaches managed to have very much faster storage periods but with the recirculating loop itself implemented in free-space \cite{Evans2023, Cheng2025, Sun2025}. Furthermore, none of these previous experiments have demonstrated higher-level functions like processing a packet with a payload containing quantum states. Therefore, an efficient, fully fibre integrated approach with fast temporal responses capable of packet handling is still missing. 

Here we cover this gap by demonstrating an optical-fibre-integrated processing buffer for packet-switched quantum networks, based on a Sagnac-loop switch \cite{alarcon2020sagnacswitch} and a fibre storage line controlled by an ultra-low-loss poled-fibre phase modulator \cite{myers1991poledsilica,kazansky1997poling,fujiwara1995uvpoling,fokine2002mzswitch, Spegel-Lexne2025}. This key device exhibits much lower losses compared with commercial fibre-optical lithium niobate (LiNbO$_3$) modulators, as well as exhibiting polarisation independence, a key property for handling polarisation-encoded quantum states.  We successfully demonstrate storage, and subsequent retrieval of a polarisation qubit payload across the fibre buffer, as well as active readout of a header while the payload remains stored. Our results demonstrate a practical and fully telecom compatible fibre architecture for processing buffers for the quantum internet. Parallel to this work in a companion submission \cite{Guerrero2026}, a completely distinct application of poled fiber modulators for quantum information was demonstrated, through the construction of an ultra-high efficiency active receiver for high-dimensional photonic quantum states in a quantum key distribution demonstration. Together we demonstrate the first uses of poled fibre phase modulators for quantum information processing applications, whose unique characteristics, namely, ultra-low-loss, fast response times, polarisation state insensitivity and direct fibre compatibility may open the path forward for highly demanding quantum communication applications. 

\section*{Results}

\subsection*{Quantum network buffering}

\begin{figure}[ht!]
    \centering
    \includegraphics[width=0.5\textwidth]{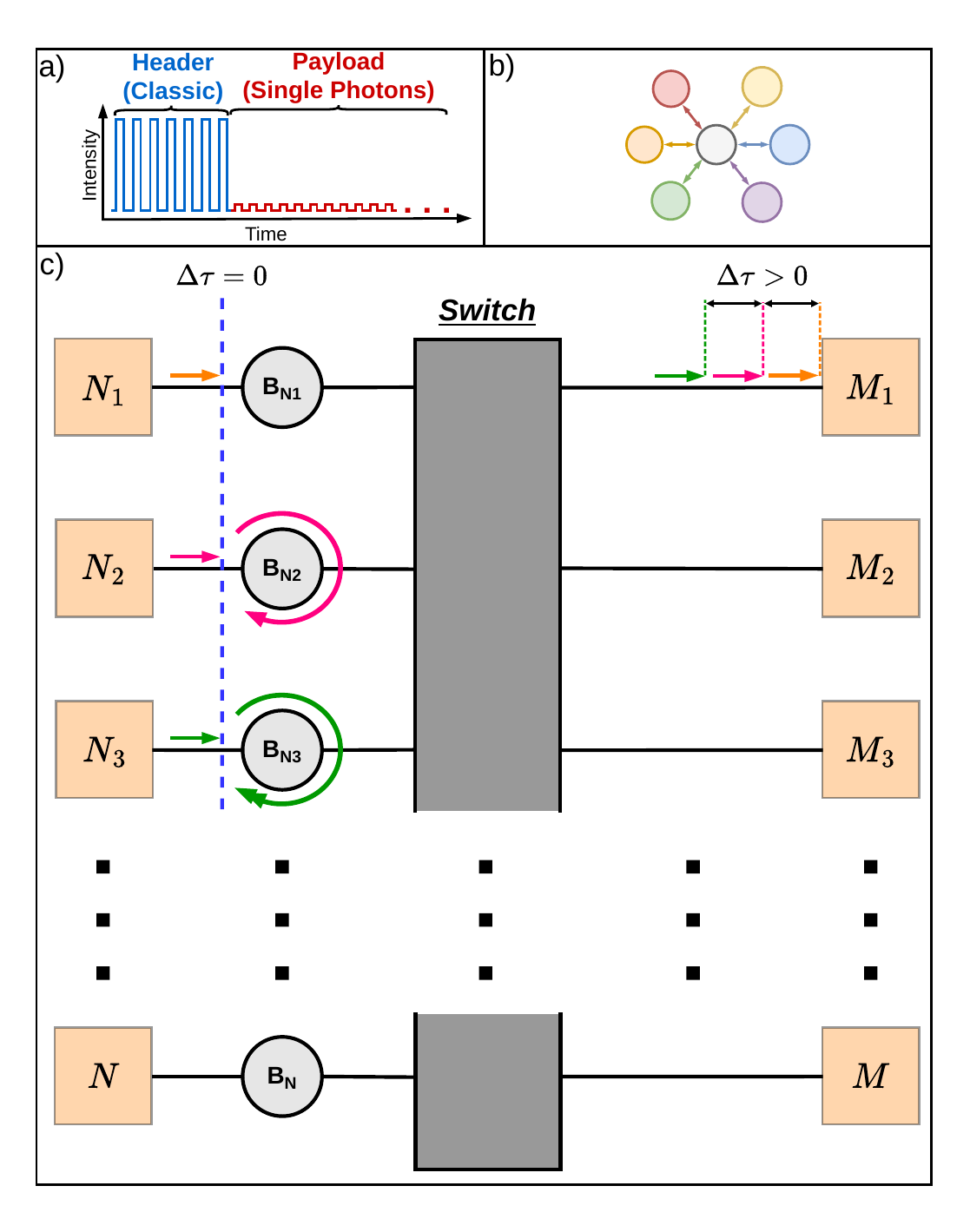}
    \caption{Quantum network packet switching. a) Packet structure consisting of a header containing classical information and a payload consisting of quantum states. b) A star network configuration, with multiple users connected through a central switching node. c) $N$ users are connected to $M$ users in a star configuration, where the central node has buffering and switching functionality. Packets can be sent from any of the $N$ to the $M$ users at will. As an example, if packets from users $N_1$, $N_2$ and $N_3$, which are all addressed to the same end user ($M_1$), arrive simultaneously at the central node, then the packets will collide in the outgoing communication channel to $M_1$. Buffering is therefore needed to time-multiplex the packets to avoid collisions in the outgoing link. The central node also needs to process the header, which contains source and destination information for instance. With this information, the delay amount in the buffer is decided by the node. Our implementation allows both header processing and buffering functionality for a packet. 
   }
    \label{fig:buffering}
\end{figure}

The quantum internet requires switching and real-time processing of packets of quantum states at the network nodes \cite{DiAdamo2022, Mandil2023, Yoo2024}, in a similar fashion to the standard internet. A typical packet structure consists of a payload containing a number of quantum states and a classical header, which holds classical information needed for real-time processing of the packet at each network node (Fig. \ref{fig:buffering}a). We encode the header and payload at the same wavelength, with our implementation being able to split the header from the payload within the buffer, owing to the high-speed response of the poled phase modulator.

A typical quantum networking scenario occurs in a star network, where multiple users are connected to a central node, responsible for processing and rerouting the packets (Fig. \ref{fig:buffering}b). As an example, $N$ simultaneous users wish to distribute quantum states to $M$ other users, all connected through the central processing node, with a multi-port switch that can connect any input to any output (Fig. \ref{fig:buffering}c). At the end of each quantum channel between the $N$ transmitting users and the central node, packet buffers are deployed to appropriately delay the packets before they are processed by the switch. The quantum states are sequentially prepared within a payload by each user and a header containing destination information (i.e. an address) is added to each payload. The header will be processed at the central node, to determine how to configure the switch for the correct routing operation. The node further checks whether the continuing communication channels to each of the $M$ destination users is available. If packets with the same final destination (i.e. $M_1$) arrive at the node with a time difference $\Delta\tau < \tau_p$, where $\tau_p$ is the time duration of the packet, then they will overlap in time in the outgoing channel, and the receiver will be unable to distinguish between them. Buffers are then used before the switch to delay the packets appropriately, such that they can all enter the same channel to $M_1$ without overlapping in time. 

\subsection*{Poled optical fibre phase modulator}

\begin{figure}[ht!]
    \centering
    \includegraphics[width=0.45\textwidth]{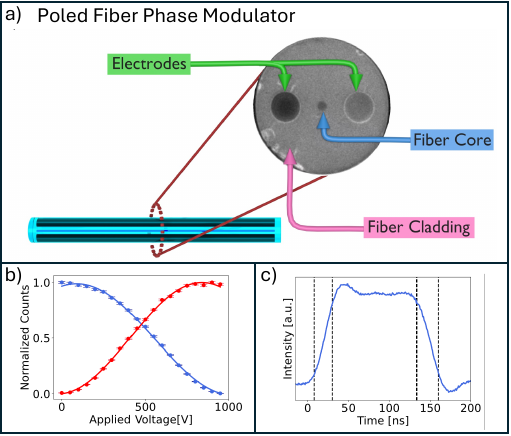}
    \caption{Poled fibre modulator. a) The cross-section of the poled fibre phase modulator is shown, imaged using a Vytran glass processor. b) Interference curves at the two outputs of a fibre-optical Sagnac interferometer driven by the poled fibre phase modulator. The outputs are measured with the two single-photon detectors. The interferometer is the same one employed in our experiment, please see in the next section for details. The error bars represent the standard deviation assuming a Poissonian photon number distribution. c) Time domain response from a driving pulse applied to the poled modulator within the Sagnac interferometer. In this measurement, the classical optical output is measured with a p-i-n photo-receiver. The rise and fall times are measured to be 22.4 ns and 26.4 ns respectively.
   }
    \label{poledmodulator}
\end{figure}

Previous active loop recirculating memories often rely on bulk electro-optical modulators for storage/retrieval control, which limits integration capabilities within optical fibre communication systems. Commercial fibre pig-tailed electro-optical modulators have typically high insertion losses ($\sim 50\%$) that greatly hamper their use for multi-pass quantum photonics applications or measurements that have a high minimum threshold detection efficiency. Therefore we resort to optical fibre poling technology to construct an in-fibre low-loss phase modulator for the recirculating loop memory. The poled fibre phase modulator used here was fabricated from a specially designed single-mode silica fibre containing two longitudinal holes running parallel to the core. These holes were drilled at the preform stage before fibre drawing and subsequently filled with bismuth electrodes in a pressure chamber, producing the cross-section shown in Fig. \ref{poledmodulator}a \cite{fokine2002mzswitch}.

 The application of an electric field between the electrodes causes a phase shift to be applied to the propagating light due to the Kerr effect. Due to the low optical non-linearity of glass, the optical fibre is first poled by coupling to its core 532 nm light from a Nd:YAG laser, while simultaneously applying a high-voltage of $\approx5$ kV, allowing electric charges to redistribute and leaving a long-lived recorded electric field in the glass after the poling field is removed \cite{Camara15}. After poling, the needed voltage to apply a $\pi$ phase shift decreases from approximately ($V_{\pi}\approx5$ kV) for 950 V to our device (Fig. \ref{poledmodulator}b), and in terms of time response, we obtain a rise and fall time of 22.4 and 26.4 ns respectively (Fig. \ref{poledmodulator}c). Since the phase modulator is essentially just an optical fibre, the losses are very low and mainly limited by the fibre connectors, which we measure to be 0.4 dB.

\subsection*{Fibre-optical active loop memory}

\begin{figure*}[ht!]
    \centering
    \includegraphics[width=1.0\textwidth]{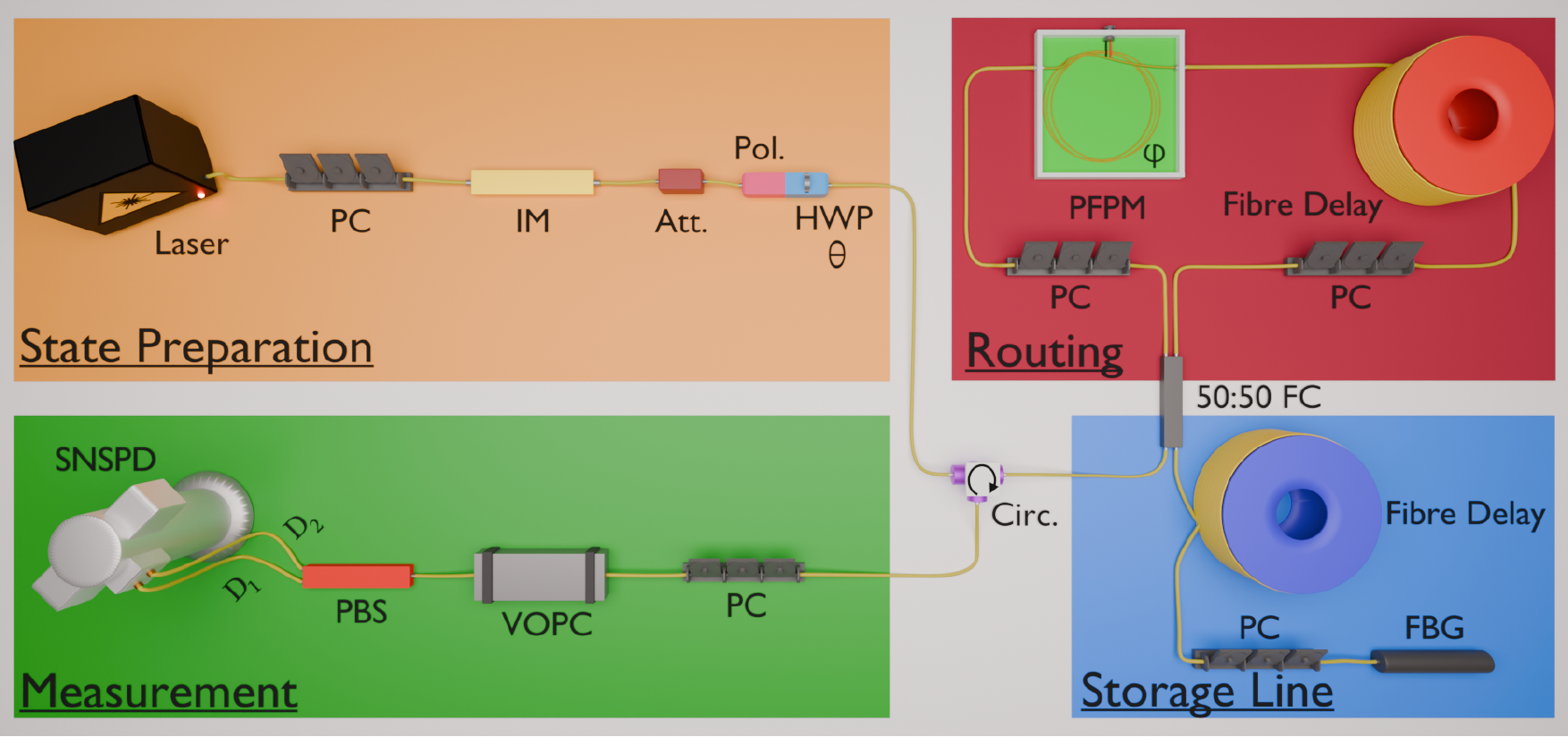}
    \caption{All-fibre quantum packet buffer implementation. We prepare weak coherent polarisation encoded qubits in the state preparation stage. A fibre coupler is employed at the output of this stage, and connected to a superconducting nanowire single-photon detector (SNSPD) for photon number calibration purposes. Then an optical circulator leads to the routing stage consisting of a Sagnac interferometer with the poled fibre phase modulator used to control whether a relative $\phi=0$ or $\phi=\pi$ phase shift is applied onto the packets, thus causing it to be routed to the storage line or to exit the loop back through the circulator. A fibre delay is employed in the Sagnac loop to ensure the phase shift is applied to the entire packet in only one of the counter-propagating directions. The storage line consists of a 100 m fibre delay, a manual polarisation controller and a mirror composed of a fibre Bragg grating. The measurement stage consists of a manual polarisation controller and a polarising beam splitter, whose two outputs are connected to two other SNSPDs. Att: Optical attenuator;  Circ: Optical circulator; FBG: Fibre Bragg grating; HWP: Half-wave plate; IM: Intensity modulator; PC: Manual polarisation controller; PFPM: Poled fibre phase modulator; Pol: Linear polariser; VOPC: Voltage operated polarisation controller; PBS: Polarising beam splitter; SNSPD: Superconducting nanowire single-photon detector.}
    \label{fig:placeholder}
\end{figure*}

A recirculating quantum memory typically consists of a resonating circuit, with an active device needed to inject the photonic quantum states onto the circuit, and then retrieve it after some chosen time \cite{Pittman2002}. We employ a combination of a fibre-optical Sagnac interferometer with a fibre delay line and a mirror to complete the resonator circuit. The Sagnac interferometer acts as an optical switch, determining whether the states are stored or retrieved. Such interferometers are very fast optical switches with several applications in quantum information \cite{Alarcon2023, Spegel-Lexne2024, Raj2025, argillander2025, Vijayadharan2025}. Our experimental setup is depicted in Fig. \ref{fig:placeholder}.

A source of weak coherent states, comprising of a continuous wave semiconductor laser diode at 1546.9 nm in tandem with an LiNbO$_3$ intensity modulator, produces a payload of 16 sequential pulses, with a period of 20 ns and a width of 10 ns at a repetition frequency of 975 Hz. A high voltage DC source feeds a high-voltage switch (Berkeley Nucleonics) capable of delivering up to 1 kV of pulse amplitude. As the modulator acts as a capacitor, it has very high impedance. Thus a 50 $\Omega$ high-voltage power resistor is connected in parallel to the modulator. This limits the repetition rate / pulse width product as the drawn current is limited by the DC source. The polarisation states of the pulses are set with a linear polariser followed by a half-wave plate (HWP). The pulses are then attenuated to the single photon level, (average 0.1 photons per pulse), approximately $1.6$ photons per payload.

The payload is then delivered to the fibre-optical Sagnac interferometer, which consists of a circulator, a 50 : 50 fibre coupler, a 1 km spooled fibre delay, the poled-fibre phase modulator and a manual polarisation controller to maximise the interference of the superposing co- and counter-propagating paths back on the fibre coupler. In one of the Sagnac outputs, the storage line is connected, consisting of a 100 m delay fibre, a manual polarisation controller and a fibre Bragg grating (FBG) acting as a mirror. The lengths of the fibre spools were chosen for convenience, and different lengths can be used depending on the desired payload length and overall storage time per cycle.

The other output, going back through the circulator, is then connected to a manual polarisation controller and a fibre polarisation beamsplitter (PBS) to perform a projective measurement on the retrieved polarisation qubits. Finally, the two outputs from the PBS are connected to two superconducting nanowire single-photon detectors (SNSPDs), within the same cryostat of the calibration detector. They have a nominal system detection efficiency $>$ 90\% and less than 30 dark counts per second for each detector. A time tagger (IdQuantique ID1000) is used to process the detection events. 

The Sagnac interferometer acts as a tunable optical switch, routing the photon between the two outputs depending on the setting $\phi$ of the poled fibre phase modulator. The relation between the applied phase difference and the probability of detecting a photon at the two outputs $D_1$ and $D_2$ of the interferometer is proportional to  $\textrm{cos}^2\left(\frac{\phi}{2}\right)$ and $\textrm{sin}^2\left(\frac{\phi}{2}\right)$ \cite{Alarcon2023}. If no relative phase is applied ($\phi=0$) the photon will exit through the same path it entered. If the phase is set to $\phi=\pi$ the photon will exit through the other output and thus, in our setup, enter the storage line. The photon is reflected back by the FBG and reenters the Sagnac loop. 

Since now the photon is entering from the opposing port, if $\phi=0$ the photon reflects back to the storage line and it becomes effectively trapped in a resonating circuit. Once a relative phase $\phi=\pi$ is applied, then the single-photon exits to the circulator and is retrieved from the memory. It then propagates to the measurement operation by the PBS before arriving at either detector $D_1$ or $D_2$.

The poled fibre phase modulator is driven by a high-voltage driver consisting of a DC power supply and a fast high-voltage switch. It generates two consecutive pulses separated in time, each 330 ns wide with 950 V amplitude, corresponding to a $\pi$ phase shift (inset Fig. \ref{poledmodulator}). The first pulse causes the qubit packet to be stored while the second retrieves it. By adjusting the delay between the two consecutive driving pulses, the photon payload can be stored and retrieved for a different number of cycles in the buffer. The period between each cycle where readout is possible is about $\approx 6 \mu s$ for the particular lengths of optical fibre in the Sagnac loop and storage line in our experiment. Note that the minimum storage time is one cycle.

\subsection*{Experimental results}

\begin{figure}[ht!]
    \centering
    \includegraphics[width=\linewidth]{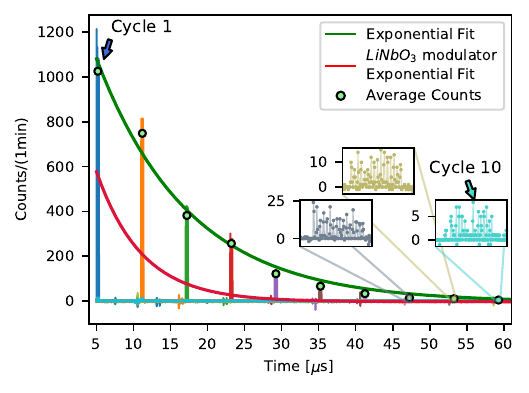}
    \caption{Buffer efficiency. A payload of 16 weak coherent states (WCSs) is stored in the buffer, and subsequently retrieved. We employ 0.1 photons per state on average (1.6 photons per payload) on the input of the buffer, to demonstrate storage for typical weak coherent state amplitudes in practical applications. The different curves corresponding up to ten storage cycles are obtained individually for each delay of the retrieval high voltage pulse, and acquired with a time tagger. They are then all plotted superposed on the figure. The green data points represent the average detected single-photon counts for all the 16 WCSs for that particular storage loop, and the green curve is an exponential decay fit. We also show for reference in the red curve, the expected performance if a commercial LiNbO$_3$ phase modulator is used instead of the poled fibre modulator. The insets zoom in on the last three storage cycles showing the individual detected counts for each WCS.}
    \label{fig:maxloopmeasurement}
\end{figure}

We initially demonstrate the retrieval of a payload over different storage cycles, as determined by the delay between the high-voltage driving pulses. The payload consisting of the train of 16 weak coherent states is sent to the input of the Sagnac loop, with the measurement data acquired by accumulating the detection events with the time tagger with each cycle delay measured with a 30 sec integration time, and all the curves are superposed and plotted in Fig. \ref{fig:maxloopmeasurement}. The PBS before the detectors was removed for this initial measurement. The manual polarisation controllers are needed to compensate residual polarisation transformations for each storage cycle, and ensure optimal interferometric visibility at the Sagnac interferometer.

We clearly see the expected exponential decay in the detection probability between storage cycles, where the losses come from the fibre spools (0.3 dB), poled phase modulator (0.4 dB), fibre coupler (0.5 dB), FBG (0.1 dB), and additional losses from connectors and splices (1.2 dB), amounting to an efficiency of $\approx 56\%$ per cycle. There is also an additional loss of approximately 1 dB for each pass through the circulator, but this only happens once independently of the number of storage cycles. We can still observe detection events following retrieval after 10 storage cycles, which corresponds to approximately 60 $\mu s$. The first storage peak in the graph corresponds to no high-voltage driving pulse to the modulator ($\phi = 0$), and is used as a reference. We also plot in Fig. \ref{fig:maxloopmeasurement} the expected performance if a commercial LiNbO$_3$ phase modulator would be used instead (typical 3 dB loss).

\begin{figure*}[ht!]
    \centering
    \includegraphics[width=\linewidth]{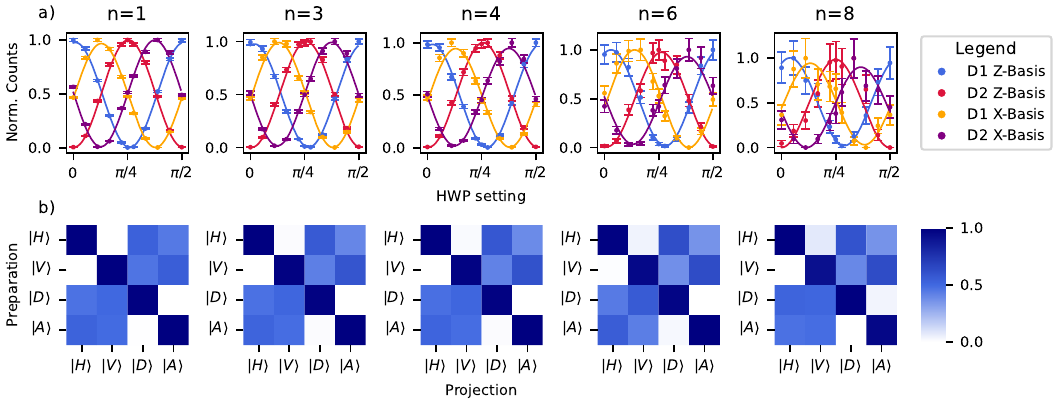}
    \caption{Quantum information storage. a) Normalised single-photon detection following the projective polarisation measurement in the measurement stage for different storage times, up to a maximum $n=8$ corresponding to 47 $\mu s$. It is carried out as a function of the input half-wave plate angle ($\theta$) in the superposition $|\psi_{\textrm{input}}\rangle = \textrm{cos}(2\theta)|H\rangle + \textrm{sin}(2\theta)|V\rangle$. For each storage cycle we project the state $|\psi_{\textrm{input}}\rangle$ over the two mutually unbiased bases used in BB84 QKD, the computational ($\mathbf{Z}$) and the logical ($\mathbf{X}$) bases. The integration time for each data point is 30 secs. Error bars come from the assumption that each detection event is governed by Poissonian statistics. b) We show the corresponding detection probabilities for each storage cycle number, by preparing each of the four prepared BB84 states and projecting onto each of the same four states, simulating a QKD protocol for different storage times, with an average QBER of $1.85 \pm 0.13\%$ over all cycles.}
    \label{fig:polstore}
\end{figure*}

We then experimentally demonstrate the storage and retrieval of polarisation-encoded weak coherent states. The PBS at the measurement stage is reinstalled, and we employ the half-wave plate $\theta$ in the beginning of the setup to generate the superposition $|\psi_{\textrm{input}}\rangle = \textrm{cos}(2\theta)|H\rangle + \textrm{sin}(2\theta)|V\rangle$, where $|H\rangle$ and $|V\rangle$ represent the horizontal and vertical polarisation states respectively. We then sweep $\theta$ and thus, for each input state, the counts at the PBS outputs are recorded over 30 s, for 5 different storage cycles, the last one corresponding to 47 $\mu$s. This experiment was done for the two mutually unbiased polarisation measurement bases used in QKD BB84-protocols \cite{bennett1984bb84}, $\mathbf{Z}$ and $\mathbf{X}$, which are changed using the manual polarisation controller before the PBS. The corresponding polarisation curves for the n$_{th}$ storage cycle are presented in figure \ref{fig:polstore}a. 

In Fig. \ref{fig:polstore}b, we show the probabilities to detect each of the four BB84 transmitted states by setting the input polarisation state angle to $\theta_{H,V} = \{0, 45\}$ and $\theta_{D,A} = \{22.5, 67.5\}$, where the subscripts $D$ and $A$ correspond to the superposition states $|D\rangle = 1 /\sqrt{2}(|H\rangle+|V\rangle)$ and $|A\rangle = 1 /\sqrt{2}(|H\rangle-|V\rangle)$. We then record the counts over the two measurement bases $\mathbf{Z}$ and $\mathbf{X}$, corresponding to the projections onto the $|H\rangle ; |V\rangle$ and $|D\rangle ; |A\rangle$ states respectively. The probability of detection for the $|i\rangle$ state is calculated as $P_{|i\rangle}=N_{D^{|i\rangle}}/(N_{D^{|i\rangle}}+N_{D^{|i\rangle\perp}})$, where $N_{D^{|i\rangle}}$ and $N_{D^{|i\rangle\perp}}$ correspond to the number of detections on the matching and on the orthogonal detector respectively, for the $|i\rangle$ input state. The quantum bit error rate (QBER) is then written as QBER = $1 - P_{|i\rangle}$, and we calculated it from the detection probabilities in Fig. \ref{fig:polstore}b for the different storage times respectively as $1.06 \pm 0.04\%$, $1.11 \pm 0.08\%$, $1.46 \pm 0.13\%$, $2.74 \pm 0.39\%$, $2.29 \pm 0.55\%$ respectively.

\begin{figure}[ht!]
    \centering
    \includegraphics[width=0.5\textwidth]{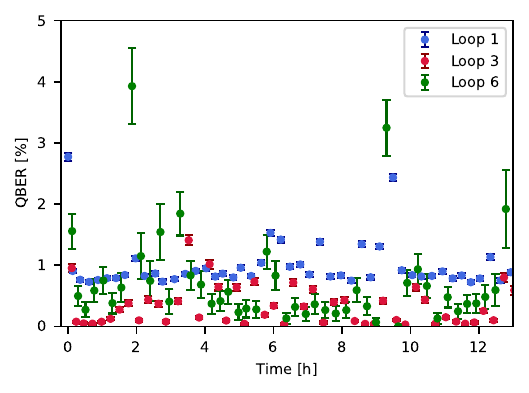}
    \caption{Long-term stability of the buffer. QBER for detecting the state $|H\rangle$ on basis $\mathbf{Z}$ as a function of time for different storage cycles. Error bars come from the assumption that each detection event is governed by Poissonian statistics.}
    \label{fig:visstab}
\end{figure}

We next evaluate the long-term stability of the buffer by continuously monitoring the QBER for a fixed input state, ($|H\rangle$). The measurement sequence starts at the first storage cycle, then proceeds to cycles n=3 and n=6, before returning to n=1. For this stability measurement, the mean input photon number is increased to 1.0, since the focus is on temporal stability. The main source of instability is the residual time-dependent birefringence of the optical fibre, which slowly changes the polarisation state reaching the measurement PBS. For each storage cycle, data is acquired for 30 s. At the start of each measurement round, the computer connected to the time tagger calculates the current detection probability (P$_{|H\rangle}$). If P$_{|H\rangle}<95\%$, corresponding to a QBER above 5\%, the measurement is paused and an automatic recalibration routine is initiated. This recalibration is performed using an all-fibre voltage-operated polarisation controller (VOPC) with 0.2 dB insertion loss, which realigns the polarisation state of the retrieved weak coherent states with the PBS before data acquisition resumes. The procedure typically takes a few minutes, depending on the initial polarisation condition. The longest storage case is the most sensitive to residual birefringence, as expected, because of the longer fibre propagation distance. Such automatic polarisation stabilisation is routinely used in fibre-based experiments and commercial QKD systems. The results, shown in Fig.~\ref{fig:visstab}, demonstrate stable long-term operation over more than 12 hours.

Finally, to demonstrate the full functionality of the buffer, we show processing of a header determining how long the payload should be stored, emulating the scenario in Fig. \ref{fig:buffering}. The header is added as three extra pulses to the beginning of the 16 weak coherent state payload, allowing the encoding of up to eight different addresses, assuming direct digital amplitude modulation. The central node then decides on the delay to be applied for each packet, depending on the destination address. In our proof-of-principle implementation, we map each header value to a different delay in terms of the storage cycle number. The header has a much higher average photon number compared to the payload, ensuring the header is successfully read out by the SNSPDs. Note that in an intermediate node in a network, simpler gated-mode avalanche diode-based single-photon detectors could be used to read out the header, or even high-sensitivity photo-receivers if the header intensity is tuned to be sufficiently high. The full packet structure is created from an arbitrary waveform generator producing the desired driving pulses to the amplitude modulator at the output of the laser. 

\begin{figure*}[ht!]
    \centering
    \includegraphics[width=1.0\linewidth]{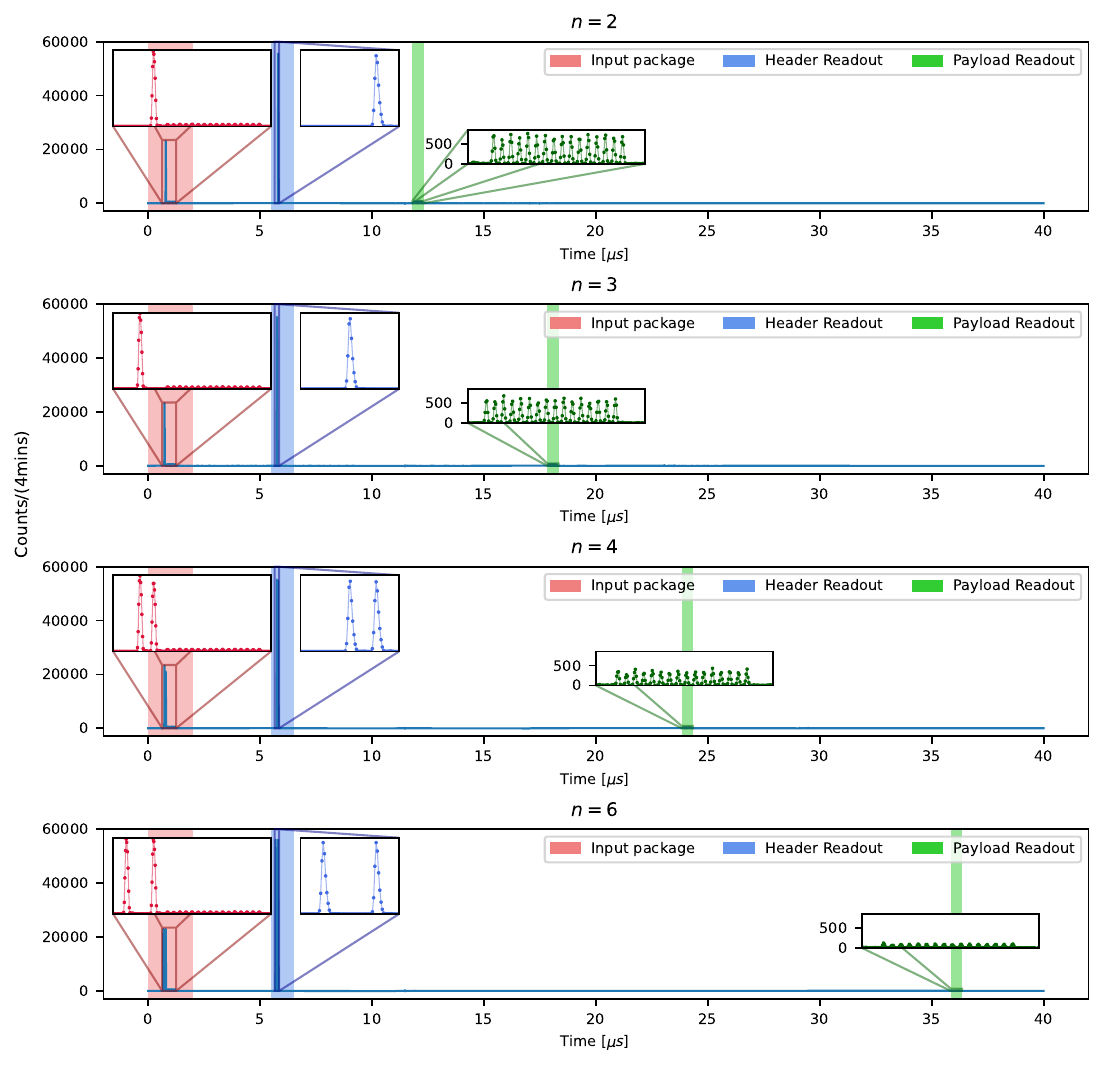}
    \caption{Quantum packet processing. The packet consists of a header containing three strong pulses, and the payload consists of 16 weak coherent states. The panels from top to bottom correspond to storage cycles of $n = \{2, 3, 4, 6\}$ respectively. On the left the red shaded waveform shows a copy of the packet before entering to the memory. The insets zoom in on the full packet before processing (red region), then the header after being split off from the payload (blue region) and the payload after being appropriately delayed (green region). Please see the text for further details. The different encoding values for the header can be seen based on on/off encoding for each respective delay for the payload on the insets.}
   \label{fig:header}
\end{figure*}

The possibility of performing real-time packet processing requires fast response times given the cumulative decoherence or loss probability as a function of storage time in a quantum memory. The poled fibre phase modulator also satisfies this strict requirement, allowing us to split the header from the payload without any need for additional time separation between the header and payload. The results are shown in Fig. \ref{fig:header}, where we see the time-of-arrival events acquired from the time-tagger for each number of different storage cycles with the corresponding header already split. We show a copy of the full packet before reaching the buffer (red shaded region), using a 10 : 90 splitter (not shown for simplicity) and the SNSPD. The header is then split off from the payload (blue shaded region), which stays in the loop and proceeds for another round. The header information containing the 3-bit binary word corresponding to the desired delay sets the delay of the applied retrieval pulse. The different retrieval times for each payload are shown in the green area. 

\begin{figure}[ht!]
    \centering
    \includegraphics[width=1.0\linewidth]{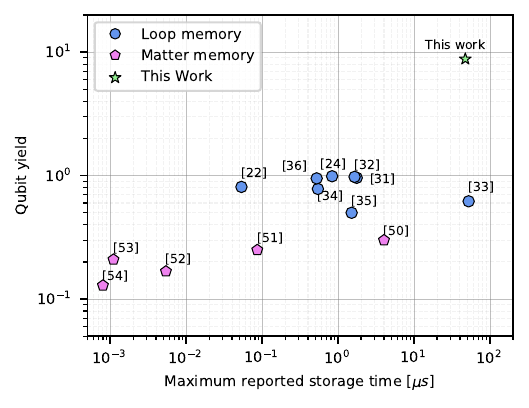}
    \caption{Effective qubit yield of room-temperature quantum  memories and photonic buffers. For each platform we plot the effective qubit yield $Y=\eta N$, which represents the expected number of qubits retrieved per storage event vs. the maximum reported storage time.  Loop memories consist of active recirculating memories while matter memories are included as a comparison, since they allow continuous on-demand retrieval. For the loop memories, we compare the different efficiencies per loop pass, while for matter memories we report the total efficiency. The efficiencies per cycle for references \cite{Bonsma-Fisher2023, Bonsma-Fisher:24, FookLee2024} were not explicitly given and were thus inferred from the experimental data provided.}
    \label{fig6}
\end{figure}

Figure~\ref{fig6} places our buffer in the broader landscape of room-temperature quantum matter memories \cite{Reim2011, Finkelstein2018, Kaczmarek2018, Thomas2023, Thomas2024} and photonic recirculating loop platforms.  We compare the effective qubit yield, which we define as $Y = \eta\,N$, where $N$ is the number of stored qubits and $\eta$ the storage efficiency per loop (or equivalently the total efficiency in the case of matter memories \cite{Reim2011, Finkelstein2018, Kaczmarek2018, Thomas2023, Thomas2024}). It quantifies the expected number of qubits recovered per storage event. In our case, through the storage and retrieval of a $16$-qubit packet at $\sim\!56\%$ per-cycle efficiency, our buffer
reaches $Y\approx 9$ qubit yield, overcoming the best previous result by nearly an order of magnitude.

\section*{Discussion}

As point-to-point quantum communication protocols get more sophisticated the focus begins to shift to networking applications. In spite of significant proof-of-principle experiments over the years to implement different networking protocols and primitives, a quantum buffering device capable of real-time packet processing is still missing. Furthermore, for such a buffer to be of more practical value, it should be easy to integrate with current telecom networking hardware.

Here we are able to demonstrate such a buffer. Our approach was done at the critical telecom spectral window, allowing full compatibility with current systems. We implement a loop memory based configuration using a Sagnac interferometer which controls the storage of the quantum states in a storage line comprised of a fibre delay and a fibre Bragg grating, operating as a mirror. A major differentiator of our implementation compared with previous demonstrations is the use of a poled fibre phase modulator, allowing our platform to be fully fibre-based with fast response times, critical for packet-processed quantum networks. Additionally, compared to typical commercial fibre pigtailed modulators, our device offers much lower insertion loss as well as low polarisation dependence, properties which have prevented previous experiments from reaching the stringent performance requirements for a practical quantum internet.  

These first results are very promising as they do not require cryogenic systems, or complex and sensitive interfaces between free-space and fibre-optics. Our proof-of-principle platform can be further improved in terms of losses, by approximately 1.1 dB per pass through the replacement of four FC/PC connectors inside the Sagnac loop with fibre splices (1.0 dB), and a fibre coupler with lower excess losses (0.1 dB), thus improving the overall transmissivity per storage cycle to 72$\%$. This value can be slightly different depending on the fibre delay values, which will be determined by the desired packet length. 

Since the buffer is designed to process entire packets, much denser payloads and headers are possible by using much narrower and closely packed pulse widths, without degrading the performance of the storage. For instance, recent results have shown the use of 45 ps wide pulses with 400 ps separation \cite{Grunenfelder2023}. The repetition rate of the buffer can be greatly improved by employing DC sources with higher current limits or a pulse generator capable of directly driving  high-impedance capacitive loads \cite{Xu2017}. Furthermore, our companion submission \cite{Guerrero2026} has demonstrated repetition rates of 100 kHz, using a poled fibre modulator with smaller V$_\pi$ and shorter pulse widths, thus confirming the potential of the device for practical quantum information processing tasks.

The results obtained are also the first to use poled fibre electro-optical modulation technology for applications in quantum information processing. The fast speed, low insertion loss and direct optical fibre compatibility make it very attractive for demanding applications such as device-independent quantum communication implementations, where low losses and high speeds for application of different measurement settings are crucial \cite{Xavier2025}. Furthermore, the polarisation independence of the device opens further possibilities, such as, in higher-dimensional hybrid entanglement processing where polarisation is one of the degrees-of-freedom \cite{Zhong2024}.

\section*{Acknowledgements}
We acknowledge helpful discussions with Alvaro Alarcón, Walter Margulis, Sebastian Etcheverry, Gustavo do Amaral and Guilherme Tempor\~ao. We acknowledge financing from Vinnova (project no. 2023-01358), the Wallenberg Center for Quantum Technologies (WACQT) and the project “DIGITAL-2022-QCI-02-DEPLOY-NATIONAL” (Project number: 101113375 - NQCIS National Quantum Communication in Sweden) funded by the European Union together with Vinnova and the Wallenberg Centre for Quantum Technology (WACQT).

\clearpage
\bibliography{sample.bib}

@article{Evans2023,
  title = {Experimental storage of photonic polarization entanglement in a broadband loop-based quantum memory},
  author = {C. J. Evans and C. M. Nunn and S. W. L. Cheng and J. D. Franson and T. B. Pittman},
  journal = {Phys. Rev. A},
  volume = {108},
  issue = {5},
  pages = {L050601},
  numpages = {6},
  year = {2023},
  month = {Nov},
  publisher = {American Physical Society},
  doi = {10.1103/PhysRevA.108.L050601},
  url = {https://link.aps.org/doi/10.1103/PhysRevA.108.L050601}
}

@article{Camara15,
author = {A. R. Camara and J. M. B. Pereira and O. Tarasenko and W. Margulis and I. C. S. Carvalho},
journal = {Opt. Express},
number = {14},
pages = {18060--18069},
publisher = {Optica Publishing Group},
title = {Optical creation and erasure of the linear electrooptical effect in silica fiber},
volume = {23},
month = {Jul},
year = {2015},
url = {https://opg.optica.org/oe/abstract.cfm?URI=oe-23-14-18060},
doi = {10.1364/OE.23.018060},
}

@article{kimble2008quantum,
  author  = {Kimble, H. J.},
  title   = {The quantum internet},
  journal = {Nature},
  year    = {2008},
  volume  = {453},
  number  = {7198},
  pages   = {1023--1030},
  doi     = {10.1038/nature07127}
}

@article{wehner2018quantum,
  author  = {Wehner, S. and Elkouss, D. and Hanson, R.},
  title   = {Quantum internet: A vision for the road ahead},
  journal = {Science},
  year    = {2018},
  volume  = {362},
  number  = {6412},
  pages   = {eaam9288},
  doi     = {10.1126/science.aam9288}
}

@article{gisin2007quantum,
  author  = {Gisin, N. and Thew, R.},
  title   = {Quantum communication},
  journal = {Nature Photonics},
  year    = {2007},
  volume  = {1},
  pages   = {165--171},
  doi     = {10.1038/nphoton.2007.22}
}

@article{pirandola2020advances,
  author  = {Pirandola, S. and Andersen, U. L. and Banchi, L. and Berta, M. and Bunandar, D. and Colbeck, R. and Englund, D. and Gehring, T. and Lupo, C. and Ottaviani, C. and Pereira,  J. L. and others},
  title   = {Advances in quantum cryptography},
  journal = {Advances in Optics and Photonics},
  year    = {2020},
  volume  = {12},
  number  = {4},
  pages   = {1012--1097},
  doi     = {10.1364/AOP.361502}
}

@inproceedings{bennett1984bb84,
  author    = {Bennett, Charles H. and Brassard, Gilles},
  title     = {Quantum cryptography: Public key distribution and coin tossing},
  booktitle = {Proceedings of the IEEE International Conference on Computers, Systems and Signal Processing},
  year      = {1984},
  pages     = {175--179},
  address   = {Bangalore, India}
}

@article{ekert1991bell,
  author  = {Ekert, Artur K.},
  title   = {Quantum cryptography based on Bell's theorem},
  journal = {Physical Review Letters},
  year    = {1991},
  volume  = {67},
  number  = {6},
  pages   = {661--663},
  doi     = {10.1103/PhysRevLett.67.661}
}

@article{lo2012mdi,
  author  = {Lo, Hoi-Kwong and Curty, Marcos and Qi, Bing},
  title   = {Measurement-Device-Independent Quantum Key Distribution},
  journal = {Physical Review Letters},
  year    = {2012},
  volume  = {108},
  number  = {13},
  pages   = {130503},
  doi     = {10.1103/PhysRevLett.108.130503}
}

@article{liao2017satelliteqkd,
  author  = {Liao, Sheng-Kai and Cai, Wen-Qi and Liu, Wei-Yue and Zhang, Liang and Li, Yang and Ren, Ji-Gang and Yin, Juan and Shen, Qi and Cao, Yuan and Li, Zheng-Ping and others},
  title   = {Satellite-to-ground quantum key distribution},
  journal = {Nature},
  year    = {2017},
  volume  = {549},
  number  = {7670},
  pages   = {43--47},
  doi     = {10.1038/nature23655}
}

@article{yin2017satelliteentanglement,
  author  = {Yin, Juan and Cao, Yuan and Li, Yu-Huai and Liao, Sheng-Kai and Zhang, Liang and Ren, Ji-Gang and Cai, Wen-Qi and Liu, Wei-Yue and Li, Bo and Dai, Hui and others},
  title   = {Satellite-based entanglement distribution over 1200 kilometers},
  journal = {Science},
  year    = {2017},
  volume  = {356},
  number  = {6343},
  pages   = {1140--1144},
  doi     = {10.1126/science.aan3211}
}

@article{lvovsky2009memory,
  author  = {Lvovsky, Alexander I. and Sanders, Barry C. and Tittel, Wolfgang},
  title   = {Optical quantum memory},
  journal = {Nature Photonics},
  year    = {2009},
  volume  = {3},
  pages   = {706--714},
  doi     = {10.1038/nphoton.2009.231}
}

@article{heshami2016memories,
  author  = {Heshami, Khabat and England, Duncan G. and Humphreys, Peter C. and Bustard, Philip J. and Acosta, Victor M. and Nunn, Joseph and Sussman, Benjamin J.},
  title   = {Quantum memories: emerging applications and recent advances},
  journal = {Journal of Modern Optics},
  year    = {2016},
  volume  = {63},
  pages   = {2005--2028},
  doi     = {10.1080/09500340.2016.1148212}
}

@article{bussieres2013applications,
  author  = {Bussi{\`e}res, F{\'e}lix and Sangouard, Nicolas and Afzelius, Mikael and de Riedmatten, Hugues and Simon, Christoph and Tittel, Wolfgang},
  title   = {Prospective applications of optical quantum memories},
  journal = {Journal of Modern Optics},
  year    = {2013},
  volume  = {60},
  pages  = {1519--1537},
  doi     = {10.1080/09500340.2013.856482}
}

@article{pompili2022networkstack,
  author  = {Pompili, Matteo and Delle Donne, Carlo and te Raa, Ingmar and van der Vecht, Bart and Skrzypczyk, Matthew and Ferreira, Guilherme and de Kluijver, Lisa and Stolk, Arian J. and Hermans, Sophie L. N. and Pawe{\l}czak, Przemys{\l}aw and Kozlowski, Wojciech and Hanson, Ronald and Wehner, Stephanie},
  title   = {Experimental demonstration of entanglement delivery using a quantum network stack},
  journal = {npj Quantum Information},
  year    = {2022},
  volume  = {8},
  pages   = {121},
  doi     = {10.1038/s41534-022-00631-2}
}

@article{alarcon2020sagnacswitch,
  author  = {Alarc{\'o}n, A. and Gonz{\'a}lez, P. and Cari{\~n}e, J. and Lima, G. and Xavier, G. B.},
  title   = {Polarization-independent single-photon switch based on a fiber-optical Sagnac interferometer for quantum communication networks},
  journal = {Optics Express},
  year    = {2020},
  volume  = {28},
  number  = {22},
  pages   = {33731},
  doi     = {10.1364/OE.408637}
}

@article{myers1991poledsilica,
  author  = {Myers, Robert A. and Mukherjee, Naren and Brueck, Steven R. J.},
  title   = {Large second-order nonlinearity in poled fused silica},
  journal = {Optics Letters},
  year    = {1991},
  volume  = {16},
  pages   = {1732--1734},
  doi     = {10.1364/OL.16.001732}
}

@article{kazansky1997poling,
  author  = {Kazansky, Peter G. and Russell, Philip St. J. and Takebe, Hiroyuki},
  title   = {Glass fiber poling and applications},
  journal = {Journal of Lightwave Technology},
  year    = {1997},
  volume  = {15},
  number  = {9},
  pages   = {1484--1493},
  doi     = {10.1109/50.618381}
}

@article{Fujiwara1995UVPoling,
  author  = {Fujiwara, T. and Wong, D. and Zhao, Y. and Fleming, S. and Poole, S. and Sceats, M.},
  title   = {Electro-optic modulation in germanosilicate fibre with UV-excited poling},
  journal = {Electronics Letters},
  volume  = {31},
  number  = {7},
  pages   = {573--575},
  year    = {1995},
  doi     = {10.1049/el:19950384}
}

@article{fokine2002mzswitch,
  author  = {Fokine, M. and Nilsson, L. E. and Claesson, {\AA}. and Berlemont, D. and Kjellberg, L. and Krummenacher, L. and Margulis, W.},
  title   = {Integrated fiber Mach--Zehnder interferometer for electro-optic switching},
  journal = {Optics Letters},
  year    = {2002},
  volume  = {27},
  number  = {18},
  pages   = {1643--1645},
  doi     = {10.1364/OL.27.001643}
}

@article{TF2018,
	author = {Lucamarini, M. and Yuan, Z. L. and Dynes, J. F. and Shields, A. J.},
	date = {2018/05/01},
	doi = {10.1038/s41586-018-0066-6},
	id = {Lucamarini2018},
	isbn = {1476-4687},
	journal = {Nature},
	number = {7705},
	pages = {400--403},
	title = {Overcoming the rate--distance limit of quantum key distribution without quantum repeaters},
	url = {https://doi.org/10.1038/s41586-018-0066-6},
	volume = {557},
	year = {2018}}

@article{Mandil2023,
	author = {Mandil, Reem and DiAdamo, Stephen and Qi, Bing and Shabani, Alireza},
	date = {2023/09/09},
	doi = {10.1038/s41534-023-00757-x},
	id = {Mandil2023},
	isbn = {2056-6387},
	journal = {npj Quantum Information},
	number = {1},
	pages = {85},
	title = {Quantum key distribution in a packet-switched network},
	url = {https://doi.org/10.1038/s41534-023-00757-x},
	volume = {9},
	year = {2023}}

@article{pittaluga2025deployed,
  title   = {Long-distance coherent quantum communications in deployed telecom networks},
  author  = {Pittaluga, Mirko and Lo, Yuen San and Brzosko, Adam and Woodward, Robert I. and
             Scalcon, Davide and Winnel, Matthew S. and Roger, Thomas and Dynes, James F. and
             Owen, Kim A. and Ju{\'a}rez, Sergio and Rydlichowski, Piotr and Vicinanza, Domenico and
             Roberts, Guy and Shields, Andrew J.},
  journal = {Nature},
  year    = {2025},
  volume  = {640},
  pages   = {911--917},
  doi     = {10.1038/s41586-025-08801-w}
}

@article{zhou2024combfield,
  title         = {{Independent Optical Frequency Combs Powered 546 km Field Test of Twin-Field Quantum Key Distribution}},
  author        = {Zhou, Lai and Lin, Jinping and Ge, Chengfang and Fan, Yuanbin and Yuan, Zhiliang and
                  Dong, Hao and Liu, Yang and Ma, Di and Chen, Jiu-Peng and Jiang, Cong and
                  Wang, Xiang-Bin and You, Li-Xing and Zhang, Qiang and Pan, Jian-Wei},
  journal       = {Physical Review Applied},
  year          = {2024},
  volume        = {22},
  number        = {6},
  pages         = {064057},
  doi           = {10.1103/PhysRevApplied.22.064057}
  
  

}

@article{DiAdamo2022,
  title = {Packet switching in quantum networks: A path to the quantum Internet},
  author = {DiAdamo, Stephen and Qi, Bing and Miller, Glen and Kompella, Ramana and Shabani, Alireza},
  journal = {Phys. Rev. Res.},
  volume = {4},
  issue = {4},
  pages = {043064},
  numpages = {17},
  year = {2022},
  month = {Oct},
  publisher = {American Physical Society},
  doi = {10.1103/PhysRevResearch.4.043064},
  url = {https://link.aps.org/doi/10.1103/PhysRevResearch.4.043064}
}

@article{Pittman2002,
  title = {Cyclical quantum memory for photonic qubits},
  author = {Pittman, T. B. and Franson, J. D.},
  journal = {Phys. Rev. A},
  volume = {66},
  issue = {6},
  pages = {062302},
  numpages = {4},
  year = {2002},
  month = {Dec},
  publisher = {American Physical Society},
  doi = {10.1103/PhysRevA.66.062302},
  url = {https://link.aps.org/doi/10.1103/PhysRevA.66.062302}
}

@article{Kaneda2017,
author = {Fumihiro Kaneda and Feihu Xu and Joseph Chapman and Paul G. Kwiat},
journal = {Optica},
number = {9},
pages = {1034--1037},
publisher = {Optica Publishing Group},
title = {Quantum-memory-assisted multi-photon generation for efficient quantum information processing},
volume = {4},
month = {Sep},
year = {2017},
url = {https://opg.optica.org/optica/abstract.cfm?URI=optica-4-9-1034},
doi = {10.1364/OPTICA.4.001034},
}

@article{Makino2016,
author = {Kenzo Makino  and Yosuke Hashimoto  and Jun-ichi Yoshikawa  and Hideaki Ohdan  and Takeshi Toyama  and Peter van Loock  and Akira Furusawa },
title = {Synchronization of optical photons for quantum information processing},
journal = {Science Advances},
volume = {2},
number = {5},
pages = {e1501772},
year = {2016},
doi = {10.1126/sciadv.1501772},
URL = {https://www.science.org/doi/abs/10.1126/sciadv.1501772}}

@article{Kaneda2019,
author = {F. Kaneda  and P. G. Kwiat },
title = {High-efficiency single-photon generation via large-scale active time multiplexing},
journal = {Science Advances},
volume = {5},
number = {10},
pages = {eaaw8586},
year = {2019},
doi = {10.1126/sciadv.aaw8586},
URL = {https://www.science.org/doi/abs/10.1126/sciadv.aaw8586}}

@article{Hou2023,
  title = {Entangled-State Time Multiplexing for Multiphoton Entanglement Generation},
  author = {Hou, Zhibo and Tang, Jun-Feng and Huang, Chang-Jiang and Huang, Yun-Feng and Xiang, Guo-Yong and Li, Chuan-Feng and Guo, Guang-Can},
  journal = {Phys. Rev. Appl.},
  volume = {19},
  issue = {1},
  pages = {L011002},
  numpages = {6},
  year = {2023},
  month = {Jan},
  publisher = {American Physical Society},
  doi = {10.1103/PhysRevApplied.19.L011002},
  url = {https://link.aps.org/doi/10.1103/PhysRevApplied.19.L011002}
}

@article{Meyer-Scott2022,
  title = {Scalable Generation of Multiphoton Entangled States by Active Feed-Forward and Multiplexing},
  author = {Meyer-Scott, Evan and Prasannan, Nidhin and Dhand, Ish and Eigner, Christof and Quiring, Viktor and Barkhofen, Sonja and Brecht, Benjamin and Plenio, Martin B. and Silberhorn, Christine},
  journal = {Phys. Rev. Lett.},
  volume = {129},
  issue = {15},
  pages = {150501},
  numpages = {6},
  year = {2022},
  month = {Oct},
  publisher = {American Physical Society},
  doi = {10.1103/PhysRevLett.129.150501},
  url = {https://link.aps.org/doi/10.1103/PhysRevLett.129.150501}
}

@article{Pang2023,
	author = {Pang, Xiao-Ling and Zhang, Chao-Ni and Dou, Jian-Peng and Li, Hang and Yang, Tian-Huai and Jin, Xian-Min},
	date = {2023/06/26},
	doi = {10.1038/s41534-023-00715-7},
	id = {Pang2023},
	isbn = {2056-6387},
	journal = {npj Quantum Information},
	number = {1},
	pages = {62},
	title = {Entangling motional atoms and an optical loop at ambient condition},
	url = {https://doi.org/10.1038/s41534-023-00715-7},
	volume = {9},
	year = {2023}}

@article{Weinbrenner2024,
  title = {Certifying the Topology of Quantum Networks: Theory and Experiment},
  author = {Weinbrenner, Lisa T. and Prasannan, Nidhin and Hansenne, Kiara and Denker, Sophia and Sperling, Jan and Brecht, Benjamin and Silberhorn, Christine and G\"uhne, Otfried},
  journal = {Phys. Rev. Lett.},
  volume = {132},
  issue = {24},
  pages = {240802},
  numpages = {7},
  year = {2024},
  month = {Jun},
  publisher = {American Physical Society},
  doi = {10.1103/PhysRevLett.132.240802},
  url = {https://link.aps.org/doi/10.1103/PhysRevLett.132.240802}
}

@article{FookLee2024,
doi = {10.1088/1367-2630/ad6703},
url = {https://doi.org/10.1088/1367-2630/ad6703},
year = {2024},
month = {aug},
publisher = {IOP Publishing},
volume = {26},
number = {8},
pages = {083011},
author = {Fook Lee, Kim and G\"{u}l, Gamze and Jim, Zhao and Kumar, Prem},
title = {Fiber loop quantum buffer for photonic qubits},
journal = {New Journal of Physics}
}

@article{Sun2025,
	author = {Sun, Ming-Shuo and Zhang, Chun-Hui and Luo, Yi-Zhen and Wang, Shuang and Liu, Yun and Li, Jian and Wang, Qin},
	doi = {10.1063/5.0255199},
	issn = {0003-6951},
	journal = {Applied Physics Letters},
	month = {03},
	number = {10},
	pages = {104001},
	title = {On-demand storing time-bin qubit states with optical quantum memory},
	url = {https://doi.org/10.1063/5.0255199},
	volume = {126},
	year = {2025}}

@article{Cheng2025,
	author = {Cheng, Sandra and Evans, Carson and Pittman, Todd},
	date = {2025/10/17},
	doi = {10.1038/s41534-025-01109-7},
	id = {Cheng2025},
	isbn = {2056-6387},
	journal = {npj Quantum Information},
	number = {1},
	pages = {163},
	title = {Fiber-coupled broadband quantum memory for polarization-encoded photonic qubits},
	url = {https://doi.org/10.1038/s41534-025-01109-7},
	volume = {11},
	year = {2025}}

@article{Bonsma-Fisher:24,
author = {K. A. G. Bonsma-Fisher and R. Tannous and D. Poitras and C. Hnatovsky and S. J. Mihailov and P. J. Bustard and D. G. England and B. J. Sussman},
journal = {Optica Quantum},
number = {1},
pages = {41--45},
publisher = {Optica Publishing Group},
title = {Storage of telecom wavelength heralded single photons in a fiber cavity quantum memory},
volume = {2},
month = {Feb},
year = {2024},
url = {https://opg.optica.org/opticaq/abstract.cfm?URI=opticaq-2-1-41},
doi = {10.1364/OPTICAQ.506601},
}

@article{Hanamura2025,
  title = {Scalable Optical Quantum State Synthesizer with Dual-Mode Resonator Memory},
  author = {Hanamura, Fumiya and Takase, Kan and Hirota, Kazuki and Nehra, Rajveer and Lang, Florian and Miki, Shigehito and Terai, Hirotaka and Yabuno, Masahiro and Kashiwazaki, Takahiro and Inoue, Asuka and Umeki, Takeshi and Asavanant, Warit and Endo, Mamoru and Yoshikawa, Jun-ichi and Furusawa, Akira},
  journal = {PRX Quantum},
  volume = {6},
  issue = {4},
  pages = {040336},
  numpages = {17},
  year = {2025},
  month = {Nov},
  publisher = {American Physical Society},
  doi = {10.1103/ztqs-q4ql},
  url = {https://link.aps.org/doi/10.1103/ztqs-q4ql}
}

@article{Alarcon2023,
author = {A. Alarc\'{o}n and J. Argillander and D. Spegel-Lexne and G. B. Xavier},
journal = {Opt. Express},
number = {6},
pages = {10673--10683},
publisher = {Optica Publishing Group},
title = {Dynamic generation of photonic spatial quantum states with an all-fiber platform},
volume = {31},
month = {Mar},
year = {2023},
url = {https://opg.optica.org/oe/abstract.cfm?URI=oe-31-6-10673},
doi = {10.1364/OE.481974},
}

@article{Spegel-Lexne2024,
author = {Daniel Spegel-Lexne  and Santiago Gómez  and Joakim Argillander  and Marcin Pawłowski  and Pedro R. Dieguez  and Alvaro Alarcón  and Guilherme B. Xavier },
title = {Experimental demonstration of the equivalence of entropic uncertainty with wave-particle duality},
journal = {Science Advances},
volume = {10},
number = {49},
pages = {eadr2007},
year = {2024},
doi = {10.1126/sciadv.adr2007},
URL = {https://www.science.org/doi/abs/10.1126/sciadv.adr2007}}

@article{Raj2025,
	author = {Raj, Chithra and Prasad, Tushita and Chaturvedi, Anubhav and Pollyceno, Lucas and Spegel-Lexne, Daniel and G{\'o}mez, Santiago and Argillander, Joakim and Alarc{\'o}n, Alvaro and Xavier, Guilherme B. and Paw{\l}owski, Marcin and Dieguez, Pedro R.},
	date = {2025/12/23},
	doi = {10.1038/s41534-025-01160-4},
	id = {Raj2025},
	isbn = {2056-6387},
	journal = {npj Quantum Information},
	number = {1},
	pages = {7},
	title = {Certifying semi-device-independent security via wave-particle duality experiments},
	url = {https://doi.org/10.1038/s41534-025-01160-4},
	volume = {12},
	year = {2025}}

@misc{argillander2025,
      title={High-dimensional detection-loophole-free measurement-device-independent quantum random number generator}, 
      author={Joakim Argillander and Daniel Spegel-Lexne and Martin Clason and Pedro R. Dieguez and Marcin Pawłowski and Anubhav Chaturvedi and Guilherme B. Xavier},
      year={2025},
      eprint={2510.06317},
      archivePrefix={arXiv},
      primaryClass={quant-ph},
      url={https://arxiv.org/abs/2510.06317}, 
}

@article{Xavier2025,
	author = {Xavier, Guilherme B. and Larsson, Jan-{\AA}ke and Villoresi, Paolo and Vallone, Giuseppe and Cabello, Ad{\'a}n},
	date = {2025/07/31},
	doi = {10.1038/s41534-025-01072-3},
	id = {Xavier2025},
	isbn = {2056-6387},
	journal = {npj Quantum Information},
	number = {1},
	pages = {129},
	title = {Energy-time and time-bin entanglement: past, present and future},
	url = {https://doi.org/10.1038/s41534-025-01072-3},
	volume = {11},
	year = {2025}}

@article{Zhong2024,
author = {Zhen-Qiu Zhong and Xiao-Hai Zhan and Jia-Lin Chen and Shuang Wang and Zhen-Qiang Yin and Jia-Qi Geng and De-Yong He and Wei Chen and Guang-Can Guo and Zheng-Fu Han},
journal = {Optica},
number = {8},
pages = {1056--1061},
publisher = {Optica Publishing Group},
title = {Hyperentanglement quantum communication over a 50 km noisy fiber channel},
volume = {11},
month = {Aug},
year = {2024},
url = {https://opg.optica.org/optica/abstract.cfm?URI=optica-11-8-1056},
doi = {10.1364/OPTICA.523955},
}

@article{Grunenfelder2023,
	author = {Gr{\"u}nenfelder, Fadri and Boaron, Alberto and Resta, Giovanni V. and Perrenoud, Matthieu and Rusca, Davide and Barreiro, Claudio and Houlmann, Rapha{\"e}l and Sax, Rebecka and Stasi, Lorenzo and El-Khoury, Sylvain and H{\"a}nggi, Esther and Bosshard, Nico and Bussi{\`e}res, F{\'e}lix and Zbinden, Hugo},
	date = {2023/05/01},
	doi = {10.1038/s41566-023-01168-2},
	id = {Gr{\"u}nenfelder2023},
	isbn = {1749-4893},
	journal = {Nature Photonics},
	number = {5},
	pages = {422--426},
	title = {Fast single-photon detectors and real-time key distillation enable high secret-key-rate quantum key distribution systems},
	url = {https://doi.org/10.1038/s41566-023-01168-2},
	volume = {17},
	year = {2023}}

@article{Reim2011,
  title = {Single-Photon-Level Quantum Memory at Room Temperature},
  author = {Reim, K. F. and Michelberger, P. and Lee, K. C. and Nunn, J. and Langford, N. K. and Walmsley, I. A.},
  journal = {Phys. Rev. Lett.},
  volume = {107},
  issue = {5},
  pages = {053603},
  numpages = {4},
  year = {2011},
  month = {Jul},
  publisher = {American Physical Society},
  doi = {10.1103/PhysRevLett.107.053603},
  url = {https://link.aps.org/doi/10.1103/PhysRevLett.107.053603}
}

@article{Finkelstein2018,
author = {Ran Finkelstein  and Eilon Poem  and Ohad Michel  and Ohr Lahad  and Ofer Firstenberg },
title = {Fast, noise-free memory for photon synchronization at room temperature},
journal = {Science Advances},
volume = {4},
number = {1},
pages = {eaap8598},
year = {2018},
doi = {10.1126/sciadv.aap8598},
URL = {https://www.science.org/doi/abs/10.1126/sciadv.aap8598}}

@article{Kaczmarek2018,
  title = {High-speed noise-free optical quantum memory},
  author = {Kaczmarek, K. T. and Ledingham, P. M. and Brecht, B. and Thomas, S. E. and Thekkadath, G. S. and Lazo-Arjona, O. and Munns, J. H. D. and Poem, E. and Feizpour, A. and Saunders, D. J. and Nunn, J. and Walmsley, I. A.},
  journal = {Phys. Rev. A},
  volume = {97},
  issue = {4},
  pages = {042316},
  numpages = {10},
  year = {2018},
  month = {Apr},
  publisher = {American Physical Society},
  doi = {10.1103/PhysRevA.97.042316},
  url = {https://link.aps.org/doi/10.1103/PhysRevA.97.042316}
}

@article{Bonsma-Fisher2023,
  title = {Fiber-integrated quantum memory for telecom light},
  author = {Bonsma-Fisher, K. A. G. and Hnatovsky, C. and Grobnic, D. and Mihailov, S. J. and Bustard, P. J. and England, D. G. and Sussman, B. J.},
  journal = {Phys. Rev. A},
  volume = {108},
  issue = {1},
  pages = {012606},
  numpages = {8},
  year = {2023},
  month = {Jul},
  publisher = {American Physical Society},
  doi = {10.1103/PhysRevA.108.012606},
  url = {https://link.aps.org/doi/10.1103/PhysRevA.108.012606}
}

@article{Thomas2023,
  title = {Single-Photon-Compatible Telecommunications-Band Quantum Memory in a Hot Atomic Gas},
  author = {Thomas, S. E. and Sagona-Stophel, S. and Schofield, Z. and Walmsley, I. A. and Ledingham, P. M.},
  journal = {Phys. Rev. Appl.},
  volume = {19},
  issue = {3},
  pages = {L031005},
  numpages = {6},
  year = {2023},
  month = {Mar},
  publisher = {American Physical Society},
  doi = {10.1103/PhysRevApplied.19.L031005},
  url = {https://link.aps.org/doi/10.1103/PhysRevApplied.19.L031005}
}

@article{Thomas2024,
author = {Sarah E. Thomas  and Lukas Wagner  and Raphael Joos  and Robert Sittig  and Cornelius Nawrath  and Paul Burdekin  and Ilse Maillette de Buy Wenniger  and Mikhael J. Rasiah  and Tobias Huber-Loyola  and Steven Sagona-Stophel  and Sven Höfling  and Michael Jetter  and Peter Michler  and Ian A. Walmsley  and Simone L. Portalupi  and Patrick M. Ledingham },
title = {Deterministic storage and retrieval of telecom light from a quantum dot single-photon source interfaced with an atomic quantum memory},
journal = {Science Advances},
volume = {10},
number = {15},
pages = {eadi7346},
year = {2024},
doi = {10.1126/sciadv.adi7346},
URL = {https://www.science.org/doi/abs/10.1126/sciadv.adi7346}}

@article{Xu2017,
	author = {Xu, Yu and Chen, Wei and Liang, Hao and Li, Yu-Huai and Liang, Fu-Tian and Shen, Qi and Liao, Sheng-Kai and Peng, Cheng-Zhi},
	doi = {10.1063/1.5006827},
	issn = {2158-3226},
	journal = {AIP Advances},
	month = {11},
	number = {11},
	pages = {115210},
	title = {Megahertz high voltage pulse generator suitable for capacitive load},
	url = {https://doi.org/10.1063/1.5006827},
	volume = {7},
	year = {2017}}

@ARTICLE{Yoo2024,
  author={Ben Yoo, S. J. and Singh, Sandeep Kumar and On, Mehmet Berkay and Gül, Gamze and Kanter, Gregory S. and Proietti, Roberto and Kumar, Prem},
  journal={IEEE Communications Magazine}, 
  title={Quantum Wrapper Networking}, 
  year={2024},
  volume={62},
  number={3},
  pages={76-81},
  doi={10.1109/MCOM.001.2300067}}

@article{Ortu2022,
  title = {Storage of photonic time-bin qubits for up to 20 ms in a rare-earth doped crystal},
  author = {Ortu, Antonio and Holz{\"a}pfel, Adrian and Etesse, Jean and Afzelius, Mikael},
  journal = {npj Quantum Information},
  volume = {8},
  pages = {29},
  year = {2022},
  doi = {10.1038/s41534-022-00541-3}
}

@article{Duranti2024,
  title = {Efficient cavity-assisted storage of photonic qubits in a solid-state quantum memory},
  author = {Duranti, Stefano and Wengerowsky, S{\"o}ren and Feldmann, Leo and Seri, Alessandro and Casabone, Bernardo and de Riedmatten, Hugues},
  journal = {Optics Express},
  volume = {32},
  number = {15},
  pages = {26884--26895},
  year = {2024},
  doi = {10.1364/OE.512318}
}

@article{Vijayadharan2025,
	author = {Vijayadharan, Kannan and Bola{\~n}os, Mat{\'\i}as Rub{\'e}n and Avesani, Marco and Vallone, Giuseppe and Villoresi, Paolo and Agnesi, Costantino},
	date = {2025/11/10},
	doi = {10.1140/epjqt/s40507-025-00428-0},
	id = {Vijayadharan2025},
	isbn = {2196-0763},
	journal = {EPJ Quantum Technology},
	number = {1},
	pages = {129},
	title = {A Sagnac-based arbitrary time-bin state encoder for quantum communication applications},
	url = {https://doi.org/10.1140/epjqt/s40507-025-00428-0},
	volume = {12},
	year = {2025}}

@inproceedings{Spegel-Lexne2025,
author = {Daniel Spegel-Lexne and J. Manoel Barbosa Pereira and Alvaro Alarc\'{o}n and Joakim Argillander and Martin Clason and {\AA}sa Claesson and Kenny Hey Tow and Walter Margulis and Guilherme B. Xavier},
booktitle = {CLEO 2025},
journal = {CLEO 2025},
pages = {FF115\_4},
publisher = {Optica Publishing Group},
title = {Storage Buffer of Polarization Quantum States Based on a Poled-Fiber Phase Modulator},
year = {2025},
url = {https://opg.optica.org/abstract.cfm?URI=CLEO_FS-2025-FF115_4},
doi = {10.1364/CLEO_FS.2025.FF115_4},
}

@article{Guerrero2026,
    author = {Guerrero, N. and Villalva, N. and dos Santos, G. H. and Salazar, C. J. and Melo, C. and Castillo, F. and Cariñe, J. and Xavier, G. B. and Gómez, E. S. and Walborn, S. P. and Pereira, J. and Saavedra, G. and Lima, G.},
    title = {Poled-fibre phase modulator for efficient high-dimensional quantum measurements},
    journal = {Submitted companion manuscript},
    year = {2026}
}

\end{document}